# Demand Response Method Considering Multiple Types of Flexible Loads in Industrial Parks

Jia Cui [a*], Mingze Gao [a], Xiaoming Zhou[b], Yang Li[c], Wei Liu [d], Jiazheng Tian [e], Ximing Zhang [a]

a The School of Electrical Engineering Shenyang University of Technology, Shenyang 110870, Liaoning Province, China; cuijiayouxiang@sut.edu.cn (C.J.); 15702415859@163.com (M.Z.);zhangsimon1108@163.com (X.M.);

b State Grid Liaoning Electric Power Supply Co, Ltd., Shenyang 110006, Liaoning Province, China; 271337328@qq.com

c The School of Electrical Engineering Northeast Electric Power University, Jilin 132012, Jilin Province, China; liyang@neepu.edu.cn

d School of Information Science and Engineering Shenyang Ligong University, Shenyang 110819, Liaoning Province, China; 251306474@qq.com

e The School of Information Science and Engineering Northeastern University, Shenyang 110819, Liaoning Province, China; 1196120094@qq.com

## Abstract

With the rapid development of the energy internet, the proportion of flexible loads in smart grid is getting much higher than before. It is highly important to model flexible loads based on demand response. Therefore, a new demand response method considering multiple flexible loads is proposed in this paper to character the integrated demand response (IDR) resources. Firstly, a physical process analytical deduction (PPAD) model is proposed to improve the classification of flexible loads in industrial parks. Scenario generation, data point augmentation, and smooth curves under various operating conditions are considered to enhance the applicability of the model. Secondly, in view of the strong volatility and poor modeling effect of Wasserstein-generative adversarial networks (WGAN), an improved WGAN-gradient penalty (IWGAN-GP) model is developed to get a faster convergence speed than traditional WGAN and generate a higher quality samples. Finally, the PPAD and IWGAN-GP models are jointly implemented to reveal the degree of correlation between flexible loads. Meanwhile, an intelligent offline database is built to deal with the impact of nonlinear factors in different response scenarios. Numerical examples have been performed with the results proving that the proposed method is significantly better than the existing technologies in reducing load modeling deviation and improving the responsiveness of park loads.

## 1. Introduction

In today's large-scale development of new energy, problems such as difficulty in peak shaving and insufficient consumption have emerged gradually. During the load peak, the traditional dispatching plan increases the output of the thermal power unit to meet the electricity demand. In addition, as the scale of the park continues to expand, various load peaks in the park will exert a lot of functional pressure on the park [1]. Demand response (DR) achieves the balance of supply and demand by reducing or delaying the power load on the demand side. At the same time, the energy system revolution with a smart grid is promoted; the informatization and intelligence of the power system have been popularized [2-3]. Demand side management (DSM) has high flexibility and great response potential. The advantages of small impact on user comfort have gained wide attention. This is an effective means to improve the power grid's ability to absorb renewable energy and lower the peak-valley difference of the power system [4-6]. With the implementation of DR, the load on the user side has gradually changed from "rigid" to "flexible" [7].

At present, flexible loads play an important role in the DR strategy. Because they have changed the time of power consumption or the size of the loads to match the power grid dispatching. Literature [8] adopts the method of interruptible loads segment compensation to realize the optimal configuration of load interruption. For power systems containing wind power, flexible loads are divided into excitation loads and interruptible loads. Although this method realizes the optimal distribution of load interruption, it does not consider the wind power consumption on the load side. Literature [9] divides flexible loads into transferable loads and reduced loads. At the same time, a dispatching model aiming at wind power consumption rate, daily load peak-to-valley difference and flexible utilization of load is constructed. Both cross-elastic and self-elastic coefficients for price-based DR are considered, literature [10] builds a chance constrained decision model considering two flexible loads. The particle swarm algorithm combined stochastic simulation method introduced to improve the system economy and absorption capacity, but the model only considers the wind power absorption of the maximum wind power fluctuation to stabilize the system, and the result is too one-sided. Literature [11] builds a grid-connected aggregate power statistical model for electric vehicle charging loads and intelligent air conditioning cooling loads to solve the problems caused by a large number of electric vehicles. It uses electric vehicles and air conditioning loads to participate in the demand side response. In the above papers, only electric vehicle or air-conditioning load is considered as a flexible load, which is not global. Literature [12] studies the content of the load control protocol. When the system is overloaded due to load fluctuations, the interrupted load is compensated for power outages, hence an optimal purchase model based on electricity prices is constructed. Literatures [13-15] consider the interconnection between different wind farms in the same area, builds a joint output model of multiple wind farms. Dynamic dispatching model on the supply and demand sides of different wind farms with dependent structures is studied, but the impact of flexible load is not considered.

In summary, the existing researches have an insufficient understanding of the requirements of multi-energy forms of integrated energy systems. There is still a need for improvement in taking into account the differences in user types on the load side. In addition, the demand side response of a single multi-energy system has its limitations in terms of time and capacity. Therefore, a demand response method is proposed considering multiple types of flexible loads.

The contributions of this paper are summarized as follows:

a) An analytical deduction model is built based on the physical process for the first time. The loads in industrial parks are divided into the high-energy-consuming industrial rotating loads, high-energy-consuming industrial heating loads, and energy storage loads. Then, they are used to analyze the load distribution of various places in the park quantitatively. A better breadth and accuracy of the expanded data would be obtained by the model. The experimental results prove that the model improves the applicability of the data.

b) The high-entropy feature information generated by the analytical deduction aggregation model is input into the WGAN-gradient penalty (WGAN-GP) for data correction. The IWGAN-GP model is built for the first time, which reduces the risk of over-fitting and under-fitting. The modeling accuracy is improved because of its faster convergence speed than traditional WGAN. Therefore, the higher quality samples are get.

c) The response contribution measurement evaluation model is proposed to build an offline decision database. The flexible load response in different scenarios is recorded and stored as a reference for the actual operation of the power grid. The experimental results show that the loads are effectively used for demand response. And the offline decision database is very beneficial for optimizing the grid dispatching strategy.

The rest of the paper is organized as follows. Section 2 presents the main flexible loads of the industrial parks which are built based on the analytical deduction model of the physical process. Section 3 describes how to build a multi-type flexible load fusion IDR model. Section 4 builds a quantitative evaluation model of response contribution and scheduling to provide theoretical support for the case analysis in Section 5. Finally, a conclusion is provided in Section 6.

## 2. Typical flexible load models for industrial parks

Based on physical process analysis, the main flexible loads in the industrial parks are divided into three types: high-energy-consuming industrial rotating loads, high-energy-consuming industrial heating loads, and storage loads.

### 2.1 High energy-consuming industrial rotating loads

The high-energy-consuming industrial rotating loads are dominated by the steel rolling process. Hence the physical power model of the steel rolling process load characteristic is presented, and the load characteristic curve is shown in Figure 1, where *R1* and *R2* represent rough rolling steps. And, *F1* to *F7* represent rolling mills.

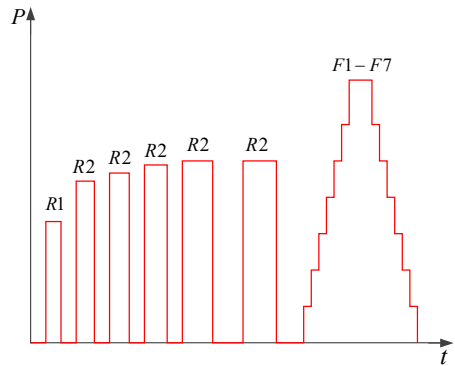

**Fig. 1.** Steel rolling process load characteristic curve

The steel rolling process load is a continuous impact load. For normal cases, the rolling mills of the steel rolling production line are divided into two types: rough rolling mills and finishing rolling mills. The rough rolling mills are reversible rolling generally. After an odd number of back-and-forth rolling, the billet passes through a single roughing mill [16]. The finishing mill is usually arranged continuously, and the billet is sent to the next process after passing through the finishing mill at one time. It is considered that the finishing mill is a roughing mill with a rolling pass of 1. After the billet passes through the rough rolling R1, it reciprocates 5 times in the rough rolling *R2* and is sent to the finishing rolling zone, then it passes through the *F1* to *F7* rolling mills at a certain speed. At this point, one billet completes a rolling process. The trapezoidal function model is used to model a single rolling mill. To use analytical expressions to reveal the law of power fluctuations of the entire steel rolling production line, the power of a single steel rolling load is simplified into a gate-shaped function whose independent variable is moment *t*, which is expressed as follows:

$$P_{roll}(t, t_0, \Delta t, a) = \begin{cases} a & t \in [t_0, t_0 + \Delta t] \\ 0 & t \notin [t_0, t_0 + \Delta t] \end{cases} \tag{1}$$

The power characteristic model of the rolling process load deduced by physical laws is described as:

$$P_{roll}^k(t) = P_{R1}^k(t) + \sum_{i=2}^{n} P_{R1}^k(t - \Delta T_{Ri}^k) + P_{F1}^k(t) + \sum_{i=2}^{n} P_{F1}^k(t - \Delta T_{Fi}^k) \tag{2}$$

where $P_{roll}^k(t)$ is the power required for *k*-type *n* billets to complete the rolling process; $P_{R1}^k(t)$ is the power required for the first billet to complete the rough rolling process; $\Delta T_{Ri}^k$ indicates the time interval between the *i*-th billet entering the rough rolling process and the first billet entering the rough rolling process; $P_{F1}^k(t)$ represents the power required to complete the finishing rolling process for the first slab and $\Delta T_{Fi}^k$ is the time interval between the *i*-th slab entering the finishing rolling process and the first slab entering the finishing rolling process [17].

### 2.2 High energy consumption industrial heating loads

High-energy-consuming industrial heating loads are dominated by smelting processes, hence the physical power model of the load characteristics of the refining furnace is represented, and the load characteristic curve is shown in Figure 2.

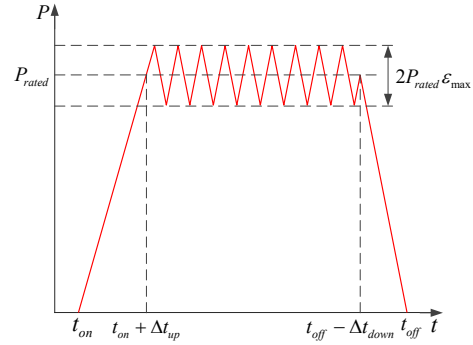

**Fig. 2.** Electric arc furnace load characteristic curve

The load of the refining furnace (a type of electric arc furnace) is an intermittent-impact load in the iron and steel industry. The main process is to melt and remove metals containing a large number of impurities such as aluminum, magnesium, and iron into a preferred liquid metal. After the power is turned on, the operator slowly lowers the electrode until the potential difference between the electrode and the charge breaks through the air to form an electric arc, and the electric arc furnace power increases rapidly within a few seconds. In the smelting process, the electric arc furnace usually runs at a certain constant gear power [18-19]. However, due to changes in the furnace temperature, charge state, and other factors, its power contains a large number of high-frequency harmonics, showing the characteristics of "band-like" power. When the electric arc furnace completes the heating task, the operator slowly lifts the electrode, and when the arc is extinguished, the power supply is interrupted. The entire shutdown process is usually completed within a few seconds. The definition function $P_{eaf}(t)$ represents the electric arc furnace power, expressed as follows:

$$P_{eaf}(t) = f\left(t, t_{on}, \Delta t_{up}, \Delta t_{down}, t_{off}, P_{rated}, \delta_{max}\right) = \begin{cases} \dfrac{P_{rated}}{\Delta t_{up}}\left(t - t_{on}\right) & t \in \left(t_{on}, t_{on} + \Delta t_{up}\right] \\ \left(1 + \varepsilon(t)\right)P_{rated} & t \in \left(t_{on} + \Delta t_{up}, t_{off} - \Delta t_{down}\right] \\ \dfrac{P_{rated}}{\Delta t_{up}}\left(t_{off} - t\right) & t \in \left(t_{off} - \Delta t_{down}, t_{off}\right] \end{cases} \quad (3)$$

where $\Delta t_{up}$ and $\Delta t_{down}$ indicate the time to rise and fall, respectively; $P_{rated}$ represents the rated power and $\delta_{max}$ is the maximum deviation. For the same electric arc furnace, $\Delta t_{up}$, $\Delta t_{down}$, $P_{rated}$, $\delta_{max}$ in formula (3) are set as constants according to the actual situation, then for an electric arc furnace, only the time series $t_{on}$ and $t_{off}$ of the start and stop within a certain period are given. That can characterize the power waveform generated by the electric arc furnace during the production process.

*2.3 Storage loads*

Most of the energy storages use electrochemical energy storage technology [20]. Lithium-ion batteries have a strong comprehensive storage capacity for energy storage and are widely used. The working principle of lithium batteries refers to the process of intercalation and intercalation of the same lithium ions and electrons inside the battery. When the battery is charged, the layered compound is decomposed to generate lithium ions electrolytically, and then moves through the electrolyte to the negative electrode to be embedded in the carbon layer. The more lithium ions are inserted, the more it is charged. When the battery is discharged, the lithium ions are removed from the carbon layer [21-22]. It falls off and then returns to the positive electrode through the movement of the electrolyte which is just the opposite of the charging process. The equivalent circuit model of the lithium battery is shown in Figure 3.

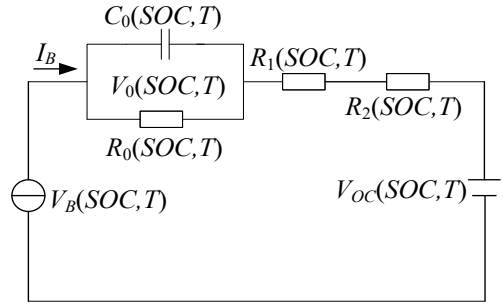

**Fig. 3.** Thevenin dynamic model of lithium battery

In Figure 3, $V_B(SOC,T)$ is the load voltage; $V_0(SOC,T)$ indicates the potential of the battery, $V_{OC}(SOC,T)$ indicates the open-circuit voltage of the battery. Equally, $R_0(SOC,T)$, $R_1(SOC,T)$ and $R_2(SOC,T)$ represent the polarization resistance, the ohmic internal resistance of the battery, and the sum of other internal resistances of the battery, respectively. $C_0(SOC,T)$ indicates the polarization capacitance and $I_B$ indicates the charge and discharge current [23]. According to the model in Figure 3, the state-space equations of the battery are built as:

$$\begin{cases} V_{OC}(SOC,T) = V_0(SOC,T) + R_0(SOC,T) \times I_B \\ + R_1(SOC,T) \times I_B + V_B(SOC,T) \\ V_0(SOC,T) = \dfrac{I_B}{C_0(SOC,T)} \times e^{-\frac{t}{R_0(SOC,T)C_0(SOC,T)}} \\ SOC = SOC_0 + \int_0^t I_B dt \end{cases} \quad (4)$$

$$\begin{cases} R_1(SOC,T) = \dfrac{\Delta V(SOC,T)}{I_B} \\ R_0(SOC,T) = \dfrac{\Delta V_1(SOC,T)}{I_B} \\ C_0(SOC,T) = \dfrac{\tau}{R_0(SOC,T)} \\ R(SOC,T) = \dfrac{V_{OC}(SOC,T) - V_B(SOC,T)}{I_B} \end{cases} \quad (5)$$

where $\Delta V$ and $\Delta V_1$ indicate the ohmic voltage and the polarization voltage respectively; $\tau$ is the polarization time constant, where $R(SOC,T)$ represents the sum of all the internal resistances of the battery. The energy storage system acts as a power source and load in the grid. When used as a peak-shaving and valley-filling dispatching application, only its charge and discharge power needs to be considered. The complex physical and chemical characteristics of its internal power electronic devices are not considered [24]. Therefore, a simple and practical power model is built. Its adjustment ability is flexible and extensive, and it can absorb and release electric energy in real-time, and participate in grid dispatching. Only its power input, output, and capacity need to be considered when scheduling.

The load characteristics have a great influence on the load modeling in the power system, and the accuracy of the load modeling determines the accuracy of the experimental data. This section obtains a load model suitable for industrial parks by classifying and modeling typical loads in industrial parks. The data is processed by the model and passed as input to the IDR model mentioned in the next section.

## 3. Multi-type flexible load fusion IDR model

The PPAD model has better interpretability. Meanwhile, the data-based inversion model method has great advantages in analyzing the problem that is difficult to be built by a causal model [25]. Therefore, the idea of using physical data fusion modeling is proposed, which is used to integrate different types of load data in industrial parks around the world.

### 3.1 Electric and heating integrated demand response

The combined electric and heating demand response of the energy system includes power demand response and heat demand response. The incentive-type power demand response cost model is the economic compensation for users participating in demand response, which is determined by the demand response agreement. Generally, this cost function is quadratic, as shown in the following formula:

$$C_{DR,i}(t) = \alpha_{DR,i}P_{DR,i}^2(t) + b_{DR,i}P_{DR,i}(t) \tag{6}$$

$$P_{DR,i,\min}(t) \le P_{DR,i}(t) \le P_{DR,i,\max}(t) \tag{7}$$

where $C_{DR,i}(t)$ represents the $i$-th incentive demand response set. The demand response cost of group users in the $t$-th period; $P_{DR,i}(t)$ represents the demand response reduction power of the $i$-th incentive-type demand-response cluster user in the $t$-th period; $\alpha_{DR,i}$ represents the quadratic coefficient of the compensation amount of the $i$-th incentive-type demand response user; $b_{DR,i}$ represents the one-time coefficient of $i$ incentive demand in response to the compensation amount of the user. Among them: $P_{DR,i,\min}(t)$ and $P_{DR,i,\max}(t)$ respectively represent the upper and lower limits of the demand response capability of the $i$-th cluster user in the $t$-th period. Other constraints for users to participate in power demand response are generally determined by specific agreements, including the limit on the number of demand responses, the shortest demand response interval, and so on.

Under the premise of ensuring comfort, the indoor temperature fluctuates within a certain interval. The indoor temperature is lowered to reduce the heat load during the peak load period. It is also increased to play a role in heat storage during the low load period. Optimizing the setting of the indoor temperature is of great significance for improving the economy of the entire integrated energy system. Based on the building's heat path model, the user's indoor temperature model is as follows:

$$T_{in}(t+1) = T_{in}(t)e^{-\Delta t/\tau} + \left[RQ_{load}(t) + T_{out}(t)\right]\left(1 - e^{-\Delta t/\tau}\right) \tag{8}$$

$$\tau = R \times C_{air} \tag{9}$$

where $T_{in}(t)$ and $T_{out}(t)$ represent the indoor and outdoor temperature respectively at the $t$ moment; $\Delta t$ represents the calculation time step; $\tau$ represents the heat dissipation time constant; $R$ represents the thermal resistance of the building; $C_{air}$ represents the indoor heat capacity; $Q_{load}(t)$ represents the heating power during the $t$ period, and $T_{in,\min}$ indicates the upper and lower limits of the room temperature respectively.

In addition, the indoor temperature should meet the comfort requirements. Therefore, the constraint conditions and the thermal load model are as follows:

$$T_{in,\min} \le T_{in}(t) \le T_{in,\max} \tag{10}$$

$$Q_{load}(t) = \frac{1}{R} \times \left[\frac{T_{in}(t+1) - T_{in}(t) \cdot e^{-\Delta t/\tau}}{1 - e^{-\Delta t/\tau}} - T_{out}(t)\right] \tag{11}$$

The cost function of heat demand response is similar to the cost function model of power demand response, both of them are quadratic functions, and will not be repeated here.

### 3.2 IDR model based on the industrial park

For the industrial park dataset, the deterministic data of the concerned issues adopts a physical simplified model to retain the causal connection between the input and output. It is input to the analytic deduction model of the physical process of the industrial parks to output high-entropy feature information, such as boundary output power. The fuzzy data is input into the inversion model based on the WGAN-GP data, thereby constructing the IWGAN-GP to improve the reliability of the analysis results.

GAN, one of the most powerful frameworks in generative models, is defined as a competitive confrontation between two interconnected neural networks. Given a noise source, the generator network tries to generate samples that are close to the real ones, while the discriminator network tries to distinguish the generated samples from the real ones. The general structure of GAN is shown in Figure 4.

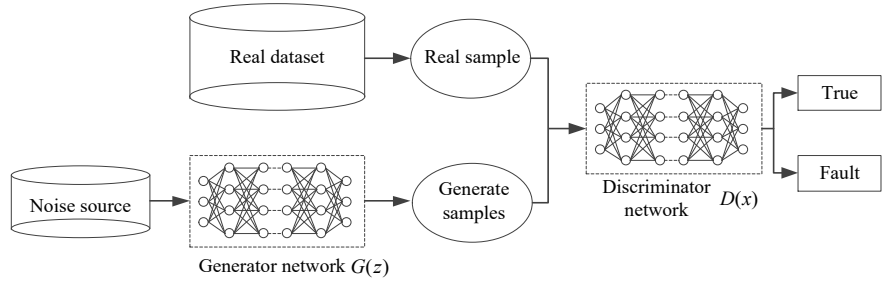

**Fig. 4.** GAN general structure

GAN consists of two models, the discriminator network $D(x)$ and the generator network $G(z)$. The essence of $D(x)$ is a differentiable function, its role is to identify the real data $x$ as much as possible, that is, the probability of $D(x)$ is 1, and the probability of $D(z)$ is 0. $G(z)$ is also a differentiable function in essence. For the randomly generated input noise sample $z$, its function is to make the noise sample imitate the real sample as much as possible, so that the discriminator network cannot distinguish between $z$ and $x$, that is, $D(x)$ should make $D(G(z))$ as 0 as possible, Instead, make $D(G(z))$ as 1 as possible.

Wasserstein-GAN (WGAN) is considered an effective alternative to traditional GAN training. By replacing JS divergence with Wasserstein distance, the current situation that JS divergence is not suitable for measuring the distance between distributions with disjoint parts is changed, and its performance is better than traditional GAN. It is very helpful for scene generation and data enhancement of missing data on different dates. However, for long-term data modeling of multi-type industrial park loads, due to its strong volatility, the modeling effect is not good. Because the Wasserstein distance needs to satisfy the continuity condition of strong Lipchitz continuity. In order to meet this condition, the traditional way is to limit the weights to a certain range to enforce the Lipchitz continuity, but the final result makes most of the weights between -0.01 and 0.01, which means that the training network is large Partial weights have only two possible numbers. For the deep neural network, the fitting ability of the deep neural network cannot be fully utilized, so that the fitting effect cannot reach the expectation. Moreover, forced clipping weights can easily lead to gradient disappearance or gradient explosion, which can easily destabilize training.

IWGAN-GP mainly improves the condition of continuity restriction than the classic WGAN. And it finds a gradient penalty (GP) way to fit the Lipchitz continuity condition. The GP establishes a loss function to make the Lipchitz restriction requirement discriminate. The gradient of the discriminator does not exceed $K$, that is, the gradient $a(D(x))$ of the discriminator is obtained first, and then the two-norm between it and $K$ is established as the loss function. However, the $D$ gradient space is the whole data sample. For data sets such as flexible loads in multi-type industrial parks, the number is too large and it is not suitable for calculation. The generated data points are randomly selected, and an interpolation is performed between the generated data and the real data, thus solving the problem of sampling over the entire data sample.

The core of the IWGAN-GP model lies in the combination mechanism of the analytical deduction model based on the physical process and the data model, using a simplified model that can reflect the physical relationship to describe the state of the industrial park load, and generating information with high entropy characteristics. As an input to the data model, it is expressed as follows:

$$\begin{cases} x_{k+1}^{'} = f(X_k) \\ x_{k+1} = h(x_{k+1}^{'}, Y_k) + u \end{cases} \tag{12}$$

where $k$ is the time state label (for example, $k+1$ represents the future moment); $f$ and $h$ reflect respectively the mapping relationship between the measured data and the features to be predicted in the analytical deduction model and data model based on the physical process; $x_{k+1}$ is the vector of system state features to be predicted at moment $k+1$; $x_{k+1}^{'}$ is the feature vector of the state to be predicted based on the analytic deduction of the physical process at moment $k+1$; $u$ is the random deviation vector in the calculation of the data model; $X_k$ and $Y_k$ are the power system measurement data compositions; the difference is that the measurement data vector $X_k$ is processed by the analytical deduction model based on the physical process, while the measurement data vector $Y_k$ is processed by the data model.

For the case of adopting the multi-type flexible load fusion model, the analytical deduction model based on the physical process provides high-entropy input features for the data model. The data model expresses the characteristics of the physical problem to be solved more accurately. This model integrates the operation data of various flexible loads and improves the accuracy of DR modeling. At the same time, it also applies to other residential and commercial electricity consumers.

## 4. Quantitative evaluation model for response contribution considering nonlinear factors

Current researches mainly focus on qualitatively evaluating the contribution of flexible loads participating in IDR. And the evaluation results are not accurate and intuitive enough. They cannot deal with the complex and changeable large power grids due to their inefficiencies of evaluation indicators to reveal the contribution of flexible loads participating in IDR. Therefore, a quantitative evaluation model of demand-side response contribution based on the improvement of RMSE index is established in this section. The model considers the black-box feature of the large power grid. The complex grid topological structure is transformed into a simple two-port network. In addition, the complex problems caused by nonlinear factors of power grid are solved by this model. It improves the current situation greatly that linear indicators cannot effectively reveal the contribution of flexible load regulation ability to participating in IDR.

### 4.1 Demand side response contribution measurement evaluation model

According to the IDR model of multiple types of flexible loads under different time-sequence operations and flexible conditions, the optimal objective function including different response indicators (abandonment wind power, peak-to-valley difference and economic efficiency) is determined, and the flexibility of each is determined. Under the constraints of the regional energy system, the simulation results of sequential operation under various flexible conditions are obtained.

For single load $X$, the demand side response contribution $\Delta Q$ is defined as:

$$\Delta Q = Q - Q'$$ 
(13)

where $Q$ is the response index quantity before response; $Q'$ is the response index quantity after response.

Due to the continuous expansion and complexity of new power grids, the nonlinear characteristics of flexible loads are prominent. It is particularly important to consider the influence of nonlinear modeling factors. To reflect the influence of nonlinear factors in the power grid, the RMSE index is optimized to the contribution model to obtain an improved demand side response supply-side contribution model $\Delta\varepsilon_x$ based on the RMSE index. The $\Delta\varepsilon_x$ is given by (14).

$$\Delta\varepsilon_x = \sqrt{\frac{1}{n}\sum_{k=1}^{n}\left(Q_k - Q_k'\right)^2}$$
(14)

where $n$ is the number of points which is determined by the step length of all-weather daily load recording; $Q_k$ is the response index of the simulated k-point running in the sequence before the response; $Q_k'$ is the response index of the simulated k-point running in the sequence after the response. The larger the value $\Delta\varepsilon_x$, the better the contribution of the flexible load in participating in the demand side response.

Similarly, for the flexible condition of multiple loads, the demand side response contribution $\Delta\varepsilon_\Omega$ based on the RMSE index improvement is:

$$\Delta\varepsilon_\Omega = \sqrt{\frac{1}{n}\sum_{k=1}^{n}\left(Q_k - Q_k'\right)^2}$$
(15)

where $\Omega = \left\{X, Y, Z, \llcorner\right\}$, which is a collection of various types of flexible loads.

### 4.2 The basic constraints of the IDR model

The intelligent offline database is built according to the IDR model of multi-type flexible load fusion. It guides flexible load dispatch of the power grid and improves the quantitative assessment of the contribution to the demand side response.

    a)    Heating loads operating power constraints

Heating loads regulation power at moment $t$ $P_{heat,t}$ should be limited between the upper and lower limits, and in line with its modeling regulation characteristics $f_{heat}$:

$$\begin{cases} P_{heat,min} \leq P_{heat,t} \leq P_{heat,max} \\ P_{heat,t} \in f_{heat}(t) \end{cases} \tag{16}$$

where $P_{heat,min}$ is the heating loads adjustment power lower limit, $P_{heat,max}$ is the heating loads adjustment power upper limit.

b) Rotating loads operating power constraints

Rotating loads regulation power at moment $t$ $P_{rot,t}$ should be limited between the upper and lower limits, and in line with its modeling regulation characteristics $f_{rot}$:

$$\begin{cases} P_{rot,min} \leq P_{rot,t} \leq P_{rot,max} \\ P_{rot,t} \in f_{rot}(t) \end{cases} \tag{17}$$

where , $P_{rot,min}$ is the heating loads adjustment power lower limit, $P_{rot,max}$ is the heating loads adjustment power upper limit.

c) Storage loads operating power constraints

$$\begin{cases} 0 \leq P_{scha,t} \leq \varepsilon_{scha,t} P_{scha,max} \\ \varepsilon_{sdis,t} P_{sdis,max} \leq P_{sdis,t} \leq 0 \\ \varepsilon_{scha,t} + \varepsilon_{sdis,t} = 1 \end{cases} \tag{18}$$

where , $P_{scha,max}$ is the upper limit of the storage power of storage devices, $P_{sdis,max}$ is the upper limit of the energy supply power of the energy storage devices (the storage power is positive and the supply power is negative), $\varepsilon_{scha,t}$ and $\varepsilon_{sdis,t}$ are equal to 0 or 1. Respectively, by introducing the symbol of energy storage and supply, the storage device can only work in one state at the same time.

d) Storage class state-of-charge constraints

$$\begin{cases} \mathrm{SOC}_t = \mathrm{SOC}_0 + \dfrac{\sum\limits_{t=1}^{T}\left(\eta_{scha}P_{scha,t} + \dfrac{1}{\eta_{sdis}}P_{sdis,t}\right)}{E_s} \\ \mathrm{SOC}_{min} \leq \mathrm{SOC}_t \leq \mathrm{SOC}_{max} \\ \mathrm{SOC}_0 = \mathrm{SOC}_T \end{cases} \tag{19}$$

where $\mathrm{SOC}_t$ is the state of charge of the storage device at moment $t$, $\mathrm{SOC}_0$ and $\mathrm{SOC}_T$ are the state of initial and final charge of the storage device, $\mathrm{SOC}_{max}$ and $\mathrm{SOC}_{min}$ are the upper and lower limits of the state of charge of the storage device, $E_s$ is the capacity of the storage device; $\eta_{scha}$ and $\eta_{sdis}$ represent the storage and energy supply efficiency of the storage type device.

e) Objective functions

$$\begin{cases} F_{pre} = \sum\limits_{i=1}^{n}\sum\limits_{t=1}^{T} Q_{pre,i,t} \\ F_{act} = \sum\limits_{i=1}^{n}\sum\limits_{t=1}^{T} Q_{pre,i,t} \\ \min F = F_{pre} - F_{act} \end{cases} \tag{20}$$

where $F_{pre}$ and $F_{act}$ represent expected response indicators and actual response indicators respectively; $Q_{pre,i,t}$ is load $i$-th amount of participation in the demand side response pre-response during period $t$, while $Q_{act,i,t}$ is load $i$-th amount of participation in the demand side response act-response during $t$ period; $\min F$ indicates the objective function of the minimum demand side response non-response index $F$.

Based on the simulation results of the multi-type flexible load fusing IDR model, the contribution degree of load participation is extracted by the quantitative evaluation model with RMSE improved demand side response [26]. The intelligent offline decision-making base is built by using RMSE to improve the quantitative assessment model of demand side response. Finally, the closed-loop structure is formed by comprehensive evaluation and optimization.

## 5. Analysis of samples

To verify the accuracy of the proposed method, the power grid flexible load data verification in Anshan City, Liaoning Province in 2019 is selected for sample verification. The collection interval is 1 hour, and there are 24 data points per day. This method could be widely applied to any region of the world and we simply use this case in China to demonstrate its implementation.

Firstly, the original data of three kinds of loads is processed by the WGAN generative confrontation network algorithm and the method of interpolation. Compared with existing methods, it is to be verified its convergence speed and quality samples. Secondly, the processed data is input into the IWGAN-GP model proposed in this paper. Through simulation, the modeling accuracy is compared with other common algorithms to deal with the risk of over-fitting and under-fitting. Finally, a response contribution measurement evaluation model and an intelligent offline database are built considering nonlinear factors. Through numerical example analysis, it is verified that the proposed demand response method acts as a great guiding role in the dispatch of flexible loads in the power grid.

### 5.1 Data processing

For heating load data, although the original data is more typical and reflects the changing trend of the flexible load power, it reflects less information, including data fluctuations, load start and stop, etc. Therefore, the interpolation method is selected to enhance the processing of single-load data. Through interpolation, the collection interval is changed from 1 hour to 15 minutes and the load data is enhanced from 24 data points (as seen in Figure 5) to 96 data points (as seen in Figure 6). Figure 6 shows that the processed data have better continuity and smoothness than Figure 5.

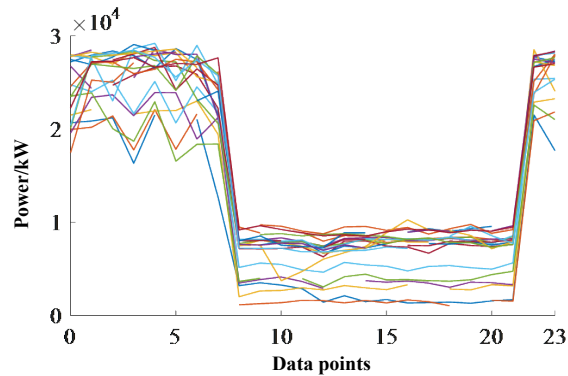

**Fig. 5.** Data without the WGAN algorithm and the interpolation method applied

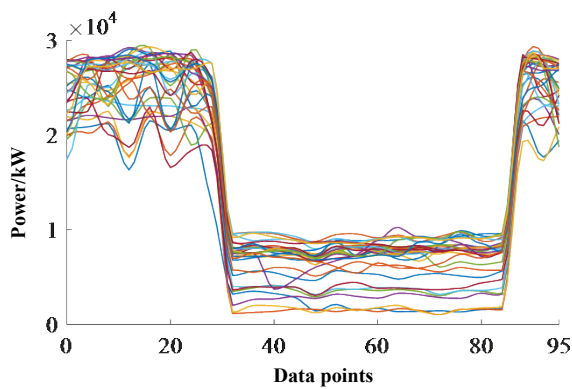

**Fig. 6.** Data with the WGAN algorithm and the interpolation method applied

For rotating load data, the figure shows that the data contained in this type of load data set are less dense and some of the data are missing (as seen in Figure 7). Therefore, it is necessary to enhance this type of data set before performing simulation analysis. This paper chooses the WGAN generative countermeasure network algorithm and the method of interpolation to enhance the original data sets. The result of a typical month's data enhancement is shown in Figure 8. The data in one month become insufficient after deleting some data due to missing data. This seriously affects the subsequent simulation analysis. However, after data enhancement, not only the collection interval is changed, but also multi-day data is generated through raw data simulation so that the data meets the simulation requirements.

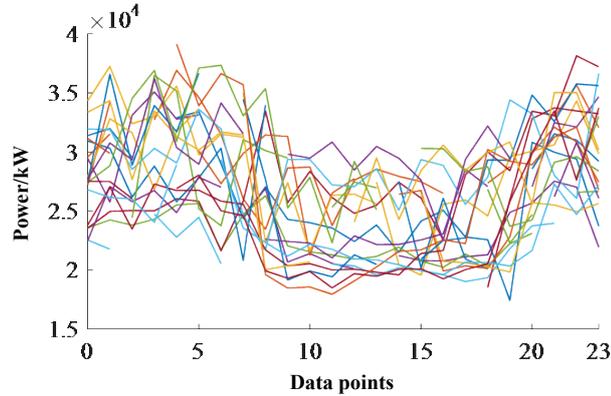

**Fig.7.** Data without the WGAN algorithm and the interpolation method applied

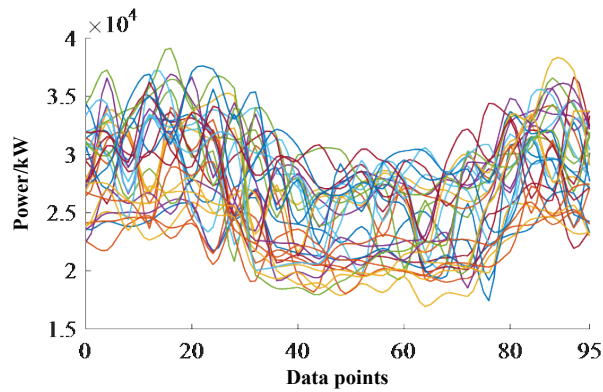

**Fig. 8.** Data with the WGAN algorithm and the interpolation method applied

*5.2 Multi-type flexible load fusion IDR model simulation results*

The pre-processed data are input into the IWGAN-GP, and the simulation results are enlarged as shown in Figure 9. It is shown that the modeling accuracy of IWGAN-GP is better than other algorithms in terms of load modeling, especially in the data peaks and valleys where there are many discrete points and large fluctuations. The Support Vector Machine (SVM) algorithm modeling results have large deviations and are affected by data fluctuations easily. Back propagation network (BP-NET) and Fourier have under-fitting phenomena, and the trend of the original data is not reflected by most of the data characteristics. In contrast, the algorithm proposed in this paper reflects the trend in the original data. The modeling effect is the best at peaks and valleys, and the modeling deviation is the smallest.

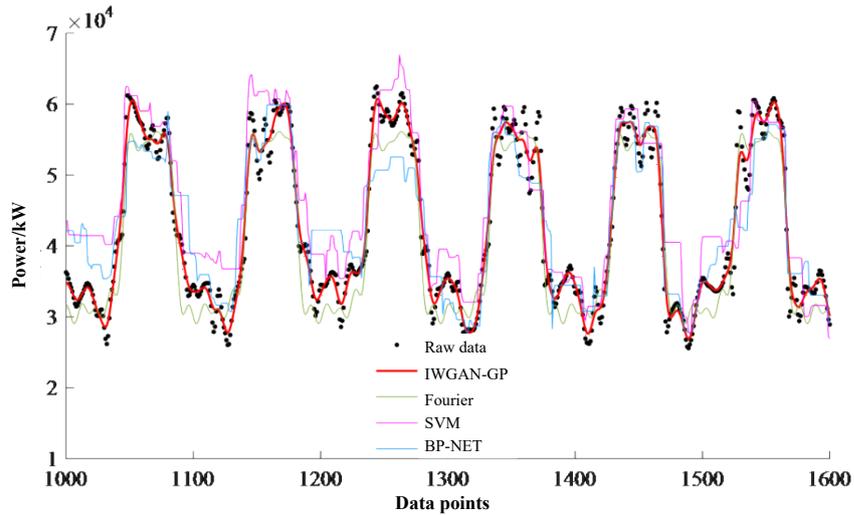

**Fig. 9.** Comparison of total monthly load modeling results in industrial parks

The monthly load modeling results show that compared with the existing load modeling algorithms, the algorithm has great advantages in long-period load data modeling. Now the steady daily loads and fluctuating daily loads of the industrial parks are used as input data into different load modeling algorithms, and the modeling results are analyzed. Figure 10 shows the total steady daily load modeling results of the industrial parks. It is seen that the curve is smooth relatively as a whole, affected by the magnitude of the power values for different types of flexible loads from the data. At the same time, the results of each load modeling method differ slightly due to fewer data points. However, from the modeling results of the part with large load fluctuations between 20-60 data points, the modeling method is better than the other three algorithms in the large data fluctuations.

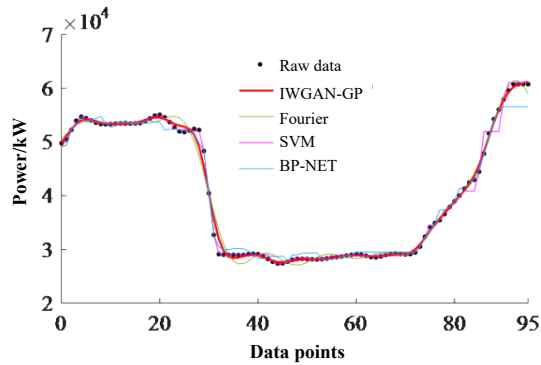

**Fig.10.** Comparison of modeling results of total steady daily load in industrial parks

Figure 11 shows the total fluctuating daily load modeling results of the industrial parks. Compared with the steady daily load data, there are more data peaks and valleys in the fluctuating daily load data, which has higher requirements for the accuracy of the load modeling method. It is seen that the performance of BP-NET for this kind of volatile load data is poor and it is inferior to the other three modeling methods in terms of modeling accuracy. Although the two load modeling methods of Fourier and SVM reflect the changing trend of fluctuating daily loads to a certain extent, the modeling effect at the peaks and valleys of the data is poor, which also deviates from the original.

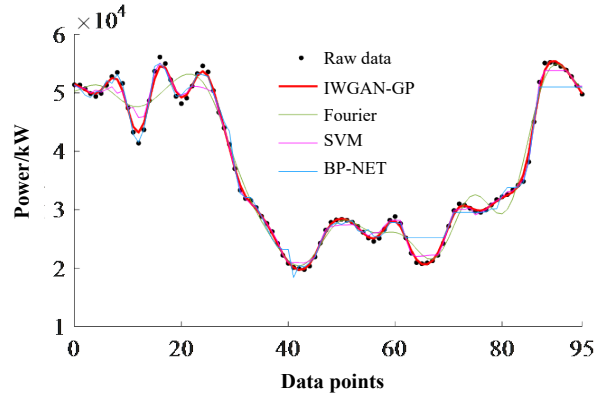

**Fig.11.** Comparison of modeling results of total fluctuating daily load in industrial parks

In summary, it is found that the modeling effect of the proposed algorithm in these three cases is better than the existing modeling method, especially in the peaks and valleys of the data. The main reason is that the modeling algorithm integrates the physical analysis process and the data inversion. The data characteristics are also retained to the greatest extent.

*5.3 Practical case comparisons of load pattern modeling*

The above load modeling results are all displayed in the form of curves, which reflect the data change trend and the difference between the modeling results of different modeling methods and the original curve during analysis. To reflect the deviations more rigorously in the modeling results, the root means square deviation (RMSD) and mean absolute deviation (MAD) are chosen for detailed comparison and analysis.

The comparison of total load modeling deviations in industrial parks is shown in table 1. It is seen from the table that IWGAN-GP has the best modeling effect when modeling monthly loads. The minimum root means square deviation is 1.813MW, which is compared with Fourier, SVM, BP-NET decreased by 2.854MW, 3.576MW, and 3.507MW respectively. At the same time, in terms of MAD, the deviation of the BP-NET modeling method is 5.392MW, which is reduced to 1.253MW. It shows the superiority of IWGAN-GP modeling performance. The IWGAN-GP modeling method is only 0.291MW when modeling steady daily load data. Compared with Fourier, SVM, and BP-NET, the MAD is reduced by 0.156MW, 0.097MW, and 0.467MW respectively, which is very close to the original data. In the results of modeling fluctuating daily load data, the effect of the IWGAN-GP modeling method is smaller than that of the steady daily load modeling results. The root means square deviations are 0.718MW and 0.582MW respectively. This is enough to reflect that the modeling method is extremely adaptable in the modeling of large volatile load data. Because it applies the physical deduction process and the mathematical inverse deduction. The problem of the existing load modeling method being affected by data fluctuations is solved to increase modeling accuracy.

**Table 1.** Comparison of total load modeling deviation in industrial parks

| Deviation | RMSD(MW) | | | | MAD(MW) | | | |
|---|---|---|---|---|---|---|---|---|
| Methods | IWGAN-GP | Fourier | SVM | BP-NET | IWGAN-GP | Fourier | SVM | BP-NET |
| Monthly load | 1.813 | 4.667 | 5.389 | 7.320 | 1.253 | 3.873 | 3.869 | 5.392 |
| Stable daily | 0.718 | 1.242 | 0.736 | 1.204 | 0.291 | 0.807 | 0.388 | 0.758 |
| Fluctuating daily | 0.582 | 2.118 | 1.131 | 1.511 | 0.409 | 1.574 | 0.680 | 0.994 |

Table 2 shows the comparison of modeling deviations of heating loads in industrial parks. It is seen from the table that IWGAN-GP has the best modeling effect when modeling heating loads. The monthly root mean square deviation of the load is 1.893MW, which is relative to Fourier, The root mean square deviation of SVM and BP-NET is significantly reduced. At the same time, in terms of MAD, compared with the BP-NET modeling method, the proposed algorithm deviation is reduced by up to 3.608MW. It shows the superiority of IWGAN-GP modeling performance. The IWGAN-GP modeling method is only 0.174MW when modeling steady daily load data. Compared with Fourier, SVM, and BP-NET, the MAD is reduced by 0.590MW, -0.035MW, and 0.323MW respectively. Among them, the small deviation from the SVM modeling method is due to the phenomenon of under-fitting in the modeling process. Although the deviation is small, the modeling result is quite different from the original data trend. In the results of modeling the heating fluctuation daily load data, the MAD of the IWGAN-GP modeling method is only 0.205MW. Compared with Fourier, SVM, and BP-NET, the MAD is reduced by 0.632MW, 0.046MW, and 0.323MW respectively. Through deviation comparison, it is seen that the modeling method has great advantages in modeling heating load data with periodic laws. Although the deviation of the SVM modeling method is small compared with this paper, it is prone to the over-fitting phenomenon of local optimal. The accuracy of the other two modeling methods is also unsatisfactory. This shows that simply using the data inversion model for load modeling has certain shortcomings, and it needs to be combined with the physical inversion model to retain the data characteristics.

**Table 2.** Comparison of heating load modeling deviations in industrial parks

| Deviation | RMSD(MW) | | | | MAD(MW) | | | |
|---|---|---|---|---|---|---|---|---|
| Methods | IWGAN-GP | Fourier | SVM | BP-NET | IWGAN-GP | Fourier | SVM | BP-NET |
| Monthly load | 1.893 | 3.633 | 5.113 | 5.925 | 1.157 | 2.915 | 3.839 | 4.765 |
| Stable daily | 0.374 | 1.093 | 0.258 | 0.858 | 0.174 | 0.764 | 0.139 | 0.497 |
| Fluctuating daily | 0.403 | 1.149 | 0.472 | 0.812 | 0.205 | 0.837 | 0.251 | 0.528 |

The comparison of modeling deviations of rotating loads in industrial parks is shown in table 3. It is seen from the table that the IWGAN-GP modeling effect is the best compared with Fourier, SVM, and BP-NET root mean square deviation when modeling monthly loads of rotating. The reduction of 2.186MW, 1.445MW, and 1.481MW is consistent with the previous analysis results. In terms of MAD, compared with the Fourier modeling method, the proposed algorithm deviation is reduced by up to 1.897MW. It shows the superiority of IWGAN-GP modeling performance. For the deviations of the modeling results of stable and fluctuating days, the deviations of the four modeling methods are reduced correspondingly due to the small number of data points and relatively few data features. The IWGAN-GP modeling method is only 0.225MW when modeling steady daily load data. Compared with Fourier, SVM, and BP-NET, the MAD is reduced by 0.415MW, 0.027MW, and 0.148MW respectively, which is very close to the original data. In the results of modeling fluctuating daily load data, the effect of the IWGAN-GP modeling method is also higher than that of the other three methods. At the same time, the rotating load data has more data features than the heating load data.

**Table 3.** Comparison of rotating load modeling deviations in industrial parks

| Deviation | RMSD(MW) | | | | MAD(MW) | | | |
|---|---|---|---|---|---|---|---|---|
| Methods | IWGAN-GP | Fourier | SVM | BP-NET | IWGAN-GP | Fourier | SVM | BP-NET |
| Monthly load | 1.167 | 3.353 | 2.612 | 2.648 | 0.840 | 2.737 | 1.955 | 1.989 |
| Stable daily | 0.299 | 0.791 | 0.484 | 0.549 | 0.225 | 0.640 | 0.252 | 0.373 |
| Fluctuating daily | 0.163 | 0.294 | 0.214 | 0.415 | 0.110 | 0.239 | 0.110 | 0.205 |

Judged from the raw load data, there is a large gap in the data of different measurement periods. The root means the square deviation is related to the magnitude. Therefore, it only reveals the superiority of the modeling method for the same measurement period in this paper. The comparison effect of the measurement period is not ideal. The goodness of fit (R-square) evaluation index is proposed to reveal the comparison effect of different measurement periods. This indicator breaks away from the magnitude constraint and reveals the comparison effect of the modeling methods of different measurement periods.

The R-square evaluation value range is between 0 and 1. The closer to 1, the better the fitting effect. Due to the small numerical range, the numerical comparison effect of different models with similar fitting effects is not ideal. Therefore, an evaluation method related to the root means square deviation and magnitude is still needed to evaluate the daily load models. Figure 12 shows the goodness of fit for the monthly loads, stable daily loads, and fluctuating daily loads modeling results of different modeling methods. Combined with the goodness of fit and the root mean square deviations, the performance of the flexible load modeling method is relatively stable. While the other three modeling methods have large differences in the goodness of fit under different conditions and the data dimension is small. The fluctuating days and stable daily loads have certain effects when modeling. But the goodness of fit drops significantly when modeling monthly loads, which is not suitable for modeling large fluctuating loads.

The monthly loads of high-energy-consuming industries have great volatility. The waveform appears as intermittent sudden gradient changes, which will cause the weight selection of the traditional neuron fitting method to change too quickly or without change. The goodness of fit is reduced greatly. The high-entropy feature information provided by the physical process analytical deduction model restricts the training process of the data inversion model. With the GP mechanism of WGAN-GP, the speed of the value change is increased and the curve regresses according to the physical rules rapidly. The fitting problems such as overshooting and local optimization are alleviated and the accuracy of modeling is improved comprehensively.

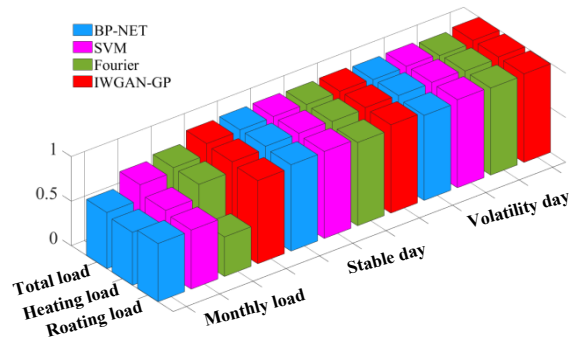

**Fig. 12.** The goodness of fit for different loads based on the deduction model

*5.4 Database building after modeling*

The modeling results need to be used as input data after completing the modeling of different types of flexible loads. The modeling result data are input into the demand side response contribution model, the quantitative evaluation model based on the RMSE index to build an intelligent offline database.

The intelligent offline database includes two indicators: response goals and flexible cases. Among them, there are three response goals: only night, all day, and only daytime. The flexible cases are mainly based on the seven kinds of combination of heating loads, rotating loads, and storage loads. The specific situation is shown in Table 4.

**Table 4.** Intelligent offline database response goals and flexible cases list

| Response target and flexible conditions | Description |
| --- | --- |
| Response goal a | Scenario 1-only night |
| Response goal b | Scenario 2-all day |
| Response goal c | Scenario 3-only daytime |
| Flexible case a | Heating loads - rotating loads - storage loads (H-R-S) |
| Flexible case b | Heating loads - rotating loads (H-R) |
| Flexible case c | Heating loads - storage loads (H-S) |
| Flexible case d | Rotating loads - storage loads (R-S) |
| Flexible case e | Storage loads (S) |
| Flexible case f | Rotating loads (R) |
| Flexible case g | Heating loads (H) |

Table 5 is the intelligent offline database built in this paper. It is seen that when the heating loads, the rotating loads, and the storage loads appear separately as flexible cases, the response values under the three response goals are relatively small, and the storage loads are different in three types. The response value under the response goals is the smallest and most stable, which is mainly determined by the characteristics of the storage flexible load. Correspondingly, the response value when the storage loads are combined with the other two loads is small relatively. The response characteristics of rotating loads are characterized by fluctuation response. This is mainly because most of the rotating loads are impactful and the working hours are irregular, so there is an all-day fluctuation response. The response value of heating loads during the day and night is greater than the response value of the whole day. This is because the working time of this type of load has a certain periodic law and responds well at night. However, when the response goal is all day, it will have a certain degree of impact on the production of such enterprises, so the response value of the heating loads in Scenario 2 is small relatively. The three flexible cases of heating-rotating, heating-storage, and rotating storage are a combination of three basic loads. Compared to a single flexible load, it is the space for scheduling when the flexible case is larger and the response value is also larger.

**Table 5.** Intelligent offline database

| Response target \ Flexible cases | H-R-S | H-R | H-S | R-S | S | R | H |
|---|---|---|---|---|---|---|---|
| Scenario a-only night(kW) | 1193.9 600 | 949.8 456 | 703.6 494 | 559.5 297 | 134.6 437 | 424.0 095 | 569.4 427 |
| Scenario b-all day(kW) | 889.2 966 | 645.5 745 | 303.8 533 | 541.4 586 | 133.6 754 | 408.0 326 | 168.7 776 |
| Scenario c-only daytime(kW) | 1472.5 340 | 1196.2 780 | 706.4 726 | 683.1 085 | 135.0 154 | 545.9 004 | 571.7 861 |

*5.5 Demand response method verification*

The scheduling model parameters are shown in table 6.

**Table 6.** Scheduling model parameters

| Parameters | Data |
|---|---|
| $P_{heat,max}$ /kW | 4500 |
| $P_{heat,min}$ /kW | 0 |
| $P_{rot,max}$ /kW | 4000 |
| $P_{rot,min}$ /kW | 0 |
| $P_{scha,max}$ /kW | 1000 |
| $P_{sdis,max}$ /kW | -1000 |
| $E_s$ /kWh | 7500 |
| $SOC_{min}$ | 0.3 |
| $SOC_{max}$ | 0.95 |
| $SOC_0$ | 0.4 |

The comparison of unresponsiveness results under various flexible conditions in scenario 1 is shown in Figure 13. It is seen from the figure that under the same scenario, the maximum unresponsiveness of storage load is much higher than that of other types of combined flexible conditions in 24 hours. On the whole, the unresponsiveness in Scenario 1 under the single-load flexible condition is significantly higher than the combined flexible condition. Among the combined flexible conditions, the R-S has the largest amount of unresponsiveness. This is mainly because the storage type and the rotation type have greater demands at night and are difficult to participate in scheduling.

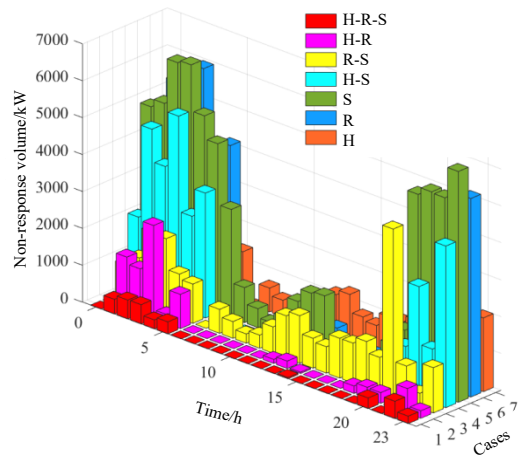

**Fig. 13.** Comparison of unresponsiveness under various flexible conditions of Scenario 1

Among all the flexible conditions, the combined flexible conditions of H-R-S have the smallest unresponse value. It means that the three loads combined dispatching has the largest response value under the same situation. Therefore it proves the effectiveness of the multi-type flexible load demand proposed in this paper.

Figure 14 shows the comparison of unresponsiveness under various flexible conditions in Scenario 2. It is seen from the figure that the non-response value of various flexible conditions in this scenario has increased compared with Scenario 1. Among them, the non-response value of the single load flexible condition is also higher than the non-response value of the combined flexible condition. In the combined flexible condition, the non-response value of the H is higher than the non-response value of the R-S and H-R. This is mainly due to the small schedulable capacity of storage loads and no connection with the other two flexible loads. In contrast, the non-response value of H-R-S in scenario 2 is the smallest, which verifies the effectiveness of the method proposed in this paper.

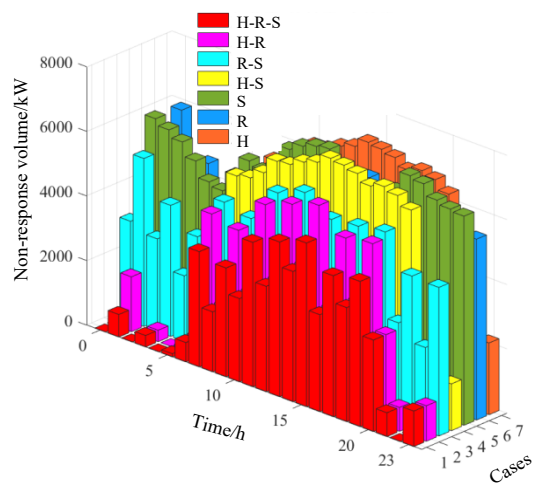

**Fig. 14.** Comparison of unresponsiveness under various flexible conditions of Scenario 2

Figure 15 shows the comparison of unresponsiveness under various flexible conditions in scenario 3. It is seen from the figure that in this scenario, the unresponsiveness values of various flexible conditions are the largest in the daytime. Among them, the non-response value of the single load flexible condition is also higher than the non-response value of the combined flexible condition. The non-response amount of heating loads in the daytime has increased significantly. Combining with its load curve, this is mainly due to the long working hours of such loads during the day and the greater load demand. In the combined flexible condition, the non-response value of the H-S is higher than the non-response value of the R-S and H-R flexible conditions. In contrast, the non-response value of H-R-S in scenario 3 is the smallest, which once again verifies the effectiveness of the method.

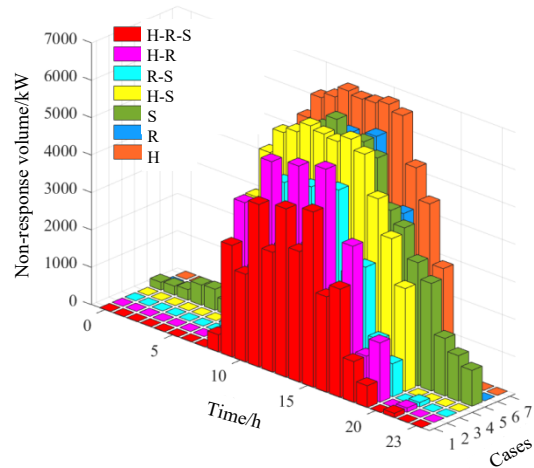

**Fig. 15.** Comparison of unresponsiveness under various flexible conditions of Scenario 3

## 6. Conclusions

A demand response method is proposed considering multiple types of flexible loads in this paper. The example analysis results show that:

a) The multi-type flexible load data of industrial parks are obtained accurately by the classification method proposed in this paper. The number of data points is tripled due to the application of interpolation. The instability of the traditional method is eliminated effectively.

b) In terms of load modeling, the modeling accuracy of the IWGAN-GP proposed in this paper is better than other algorithms significantly. The modeling results have larger deviations than the compared algorithms such as SVM. And they are easily affected by data fluctuations, especially for the cases which conclude different original data. Through the error analysis, it can be seen that the modeling effect of IWGAN-GP is the best in long-term load modeling. Compared with Fourier, SVM, and BP-NET, the root mean square error is reduced by 2.854MW, 3.576MW, and 3.507MW, respectively.

c) Flexible loads are efficiently utilized to participate in demand-side response due to the established intelligent offline database. The amount of unresponsiveness under various flexibility conditions is predicted accurately. The relative error of the amount of unresponsiveness has been reduced significantly by over 25% compared to the traditional methods.

The future work plan is to carry out detailed study on the flexible loads scheduling of industrial parks in power grid. It is also interesting to incorporate the developed load modeling method for scheduling of or community integrated energy systems.

## Acknowledgments:

This work is supported in part by the Natural Science Foundation of Liaoning Province. (20170520318) .